\documentstyle[preprint,aps]{revtex}
\tightenlines

\newcommand {\be}{\begin{equation}}
\newcommand {\ee}{\end{equation}}
\newcommand{\bey}{\begin{eqnarray}}
\newcommand{\eey}{\end{eqnarray}}

\begin{document}
%\draft
\title{Finite-length Lyapunov exponents and conductance for quasi-1D
disordered solids
}
\author{T. Kottos$^{1}$, F.M. Izrailev$^{2,3}$ and A. Politi$^4$
\\
%\address{
$^1$ Department of Physics of Complex Systems,
The Weizmann Institute of Science, Rehovot 76100, Israel\\
$^2$ Instituto de Fisica, Universidad Autonoma de Puebla,
Apdo. Postal J-48\\ 
Col. San Manuel, Puebla, 72570 Mexico\\
$^3$ Budker Institute of Nuclear Physics, 630090
Novosibirsk, Russia \\
$^4$ Istituto Nazionale di Ottica, Largo E. Fermi 6, 50125 Firenze
and INFN Sezione di Firenze, Italy
}

\date{\today}
\maketitle

\begin{abstract}
\begin{center}
\parbox{14cm}
{The transfer matrix method is applied to finite quasi-1D disordered samples
attached to perfect leads. The model is described by structured band
matrices with random and regular entries. We investigate numerically the
level spacing distribution for finite-length Lyapunov exponents as well as
the conductance and its fluctuations for different channel numbers and
sample sizes. A comparison is made with theoretical predictions and with
numerical results recently obtained with the scattering matrix approach. 
The role of the coupling and finite size effects is also discussed.}
\end{center}
\end{abstract}

%\hspace{1.9cm}
%PACS numbers: 05.45.+b, 71.55J}

\newpage

\section{\bf Introduction}

Since the discovery of non-self-averaging conductance fluctuations in
mesoscopic conductors \cite{A85,LS85,WW86}, it has become necessary to
reconsider various fundamental aspects of quantum electronic transport. In
particular, the one-parameter scaling hypothesis \cite{AALR79} has been
challenged \cite{AKL86} on the ground that the conductance $G$ is not a
self-averaging quantity, and thus it is necessary to consider its entire
distribution rather than just the mean value. One of the approaches that
provide the possibility of obtaining the conductance distribution is based
on the theory of random matrices (for a recent review see, for example, \cite
{B97}). This theory exploits the connection between the conductance and
finite-length Lyapunov exponents \cite{P84} of the transmission matrices
(see also \cite{M95} and references therein). Thus, in this approach, the
main interest is in the properties of the Lyapunov spectra (all Lyapunov
exponents ordered in an increasing/decreasing way).

In this paper we study quasi-1D disordered samples of finite size described
by Band Random Matrices (BRM). Such matrices have been extensively
investigated in applications to thin wires (see the review \cite{FM94} and
references therein) or, equivalently, to 1D Anderson-type systems with
long-range random hopping. Scaling properties of the localization length of
eigenfunctions are already well understood since some years \cite{CMI90,FM94}%
. More recently, the scaling properties of averaged Lyapunov exponents have
been also established both for infinite \cite{KPIR96} and finite \cite{KPI97}
BRMs. In this paper, we are mainly interested in statistical properties of
the Lyapunov exponents and thereby of the conductance as a function of the
sample size and the number of transmission channels.

One of the results of this paper concerns the comparison with the
predictions of the Random Matrix Theory (RMT). It is known that, in contrast
to other characteristics like the spectral rigidity, the shape of the level
spacing distribution $P(s)$ is quite insensitive to deviations from the
Wigner-Dyson surmise, i.e. the prediction of the RMT. Nevertheless, in spite
of an overall good correspondence between the $P(s)$ numerically determined
and the theoretical expression, here we show that the normalized second
moment of the distribution $P(s)$ reveals clear deviation from the
theoretical predictions when applying the RMT to the Lyapunov Spectra.
Accordingly, this deviation can serve as a sensitive measure of the
correspondence of the data to the RMT results. We attribute such a
discrepancy to the existence of a non-random repulsion acting on the low
Lyapunov exponents from the left side, due to the symmetry of the Lyapunov
spectrum. Though the observed deviation is relatively small, it manifests
itself in the value of the variance for the conductance fluctuations which
we have studied numerically by making use of the Landauer formula expressed
in terms of the Lyapunov spectra.

The outline of the paper is as follows. In the next Sec.~II we describe the
physical setup which permits us to study the properties of conductance of
finite samples attached to the leads. First, we introduce the model in terms
of Hamiltonian band matrices and derive the corresponding transfer matrices.
Using these matrices, one can compute the conductance of quasi-1D finite
samples through the Landauer formula. This is done by introducing a specific
matrix, the eigenvalues of which represent the key ingredient of the
conductance expression. At the beginning of Sec.~III, we give a very short
outlook of theoretical results about the connection between fluctuations of
the conductance and of finite-length Lyapunov exponents. According to these
findings, the statistical properties of Lyapunov exponents may be
approximately described by the RMT. In particular, the ``level'' spacing
distribution is expected to follow the famous Wigner-Dyson expression. In
this Section we compare these predictions with our numerical data obtained
for the whole Lyapunov spectrum, finding clear deviations not only in the
upper but also in the lower part of the spectrum, where a good
correspondence with RMT was expected instead. A study of both the
conductance and its fluctuations in different regimes, ranging from the very
localized to the metallic one, is performed in Sec.~IV. In particular, we
have determined the dependence of the average conductance on the related
scaling parameter when passing from the localized to metallic regime.
Another quantity of interest was the universal conductance fluctuations for
which we have checked the theoretical predictions. Detailed discussion of
our results is presented in the last Section.

\section{\bf The physical setup}

The simplest general model in the class of quasi-1D or 1D systems with long
range hopping is represented by the Schr\"odinger equation with interactions
described by Band Random Matrices. This ensemble is defined as a set of real
symmetric matrices the entries of which are independent Gaussian variables
with zero average and variance $\sigma_0 ^2=1+\delta _{n,m}$ ($\delta _{n,m}$
is the Kronecker symbol) if $|n-m|\leq b$, and zero otherwise (i.e., there
are $2b+1$ nonzero elements in any row). The parameter $b$ defines the
hopping range between neighboring sites; in the quasi-1D interpretation, it
is the number of transverse channels along a thin wire. In the study of the
electronic conduction in disordered wires at zero temperature, one uses
mainly one of the two theoretical approaches based on ideas of Thouless and
Landauer respectively. In the Thouless approach, the wire is regarded as a
closed system, and its conductance is defined through the sensitivity of
eigenvalues to changes of the boundary conditions \cite{T74,AM92,CGIMZ94}.
In the Landauer approach, one has to embed a disordered sample into a
perfectly ordered lattice, i.e. to add perfect leads both to the left and to
the right of the disordered sample. Thus, in contrast with the former
approach, the latter one deals with an open system (for a comparison of the
Thouless and Landauer conductance see, e.g., \cite{CGM97,BHMM96}).

Our model is described by the time-dependent Schr\"odinger equation, 
\begin{equation}
\label{eqmo}i\,\frac{dc_n(t)}{dt}=\,\sum_{m=n-b}^{n+b}H_{n,m}c_m\quad , 
\end{equation}
where $c_n(t)$ is the probability amplitude for an electron to be at site $n$
and $H_{n,m}$ is a symmetric banded matrix with random entries for the
disordered part (see below). The eigenvalues can be obtained by substituting
the relation $c_n(t)=\exp (-iEt)\psi _n$ in Eq.~(\ref{eqmo}) and solving the
resulting equation for $\psi _n$, 
\begin{equation}
\label{map}\psi _{n+b}=\frac 1{H_{n,n+b}}\left( E\psi
_n-\sum_{m=n-b}^{n+b-1}H_{n,m}\psi _m\right) \quad . 
\end{equation}
By defining $x_n(i)\equiv \psi _{n+b-i}$, the above equation can be recasted
in the form of a $2b$-dimensional linear map $T_n$, 
\begin{eqnarray}
\label{eq:mmap}
&& x_{n+1}(1) = \frac{1}{H_{n,n+b}} \left( E x_n(b) -
   \sum_{j=1}^{2b} H_{n,n+b-j}x_n(j) \right) \nonumber \\
&& x_{n+1}(j) = x_{n}(j+1)  \quad \quad 1 < j \le 2b
\end{eqnarray}
which provides an alternative interpretation of the stationary Schr\"odinger
equation. In this picture, an eigenstate of Eq.~(\ref{eqmo}) can be treated
as a ``trajectory'' evolving under the action of map (\ref{eq:mmap}), and
its localization properties are determined by the Lyapunov exponents. In the
previous paper \cite{KPIR96}, we have investigated the shape of the Lyapunov
spectrum in the limit of infinitely extended disordered samples. Here,
aiming at a more complete understanding of conductance properties, we choose
to rely on the multichannel Landauer formula \cite{P84} for the conductance
of finite samples. The physical set-up requires considering a disordered
sample of length $L$ in between two perfect leads. At variance with the
standard Anderson model, where only nearest-neighbor couplings are included,
here the long-range hopping terms allow some freedom in the structure of the
ordered leads and, especially, in the connection of the leads with the
disordered sample. As for the leads, a natural way is to assume a band
structure in the ordered part (leads) with $H_{n,m}=V$, for $|n|,|m|>L$ (for
the sake of simplicity we put $V=1$). As for coupling matrix elements, we
choose them in the same way as in the bulk, (random Gaussian entries with
zero mean and variance $\sigma _0^2=1$).

There is an immediate analogy with the Anderson problem on a stripe of width 
$b$, where one deals with $2b\times 2b$ matrices, too. However, in our case
the one-step matrix $T_n$ defining map (\ref{eq:mmap}) is not symplectic and
the determinant is not equal to 1, but to $%
(-1)^{2b+1}(-H_{b+1,1}/H_{b+1,2b+1})$. Yet, the total transfer matrix $%
T_N=\prod_{n=1}^NT_n$, where $N=L+2b$, satisfies the following relation 
\begin{equation}
\label{symp1}(T_N)^{\dagger }\Sigma T_N=\Sigma 
\end{equation}
where $T_N^{\dagger }$ denotes the Hermitian conjugate of $T_N$ and the
matrix $\Sigma $ is defined as 
\begin{equation}
\label{symp2}\Sigma =\\\left( 
\begin{array}{cc}
0 & S \\ 
-S^t & 0 
\end{array}
\right) . 
\end{equation}
Here $S$ is a lower triangular matrix such that $S_{ij}=1$, for any $1\le
j\le i\le b$ and $S^t$ denotes the transposed operator. In fact, Eq.~(\ref
{symp1}) means that $T_N$ has a ``generalized symplectic'' structure \cite
{M97}.

The scattering properties of the sample can be better investigated by
choosing an appropriate base, namely by decomposing the eigenfunctions in
the plane waves, 
\begin{equation}
\label{free1}\psi _n={\frac{e^{inp}}{\sqrt{2\pi }}}\;, 
\end{equation}
supported by the ordered lattice. The corresponding eigenvalues are 
\begin{equation}
\label{free2}E=1+2\cos p+...+2\cos (bp)=\frac{\sin ({\frac{2b+1}2}p)}{\sin
(\frac p2)}\;, 
\end{equation}
while the velocity of a wave packet centered around $p$ is 
\begin{equation}
\label{vel}v(p)=\frac{dE}{dp}=\frac{2b+1}2{\frac{\cos (\frac{2b+1}2p)}{\sin
(\frac p2)}}-\frac{\sin (\frac{2b+1}2p)}{2\sin {}^2(\frac p2)}\cos (p/2)\;. 
\end{equation}
For simplicity, in what follows, we study the case $E=0$ for which all
transmission channels are open. Therefore, there are $b$ pairs of open
channels with opposite velocities, 
\begin{equation}
\label{vel1}p_k=(-1)^k\frac{2\pi k}{2b+1}\quad ,\,\quad \,1\le |k|\le b\;. 
\end{equation}
The matrix $M$ connecting the amplitudes to the left with those to the right
of the scatterer is 
\begin{equation}
\label{tran}\left( 
\begin{array}{c}
{\bf A^R} \\ {\bf B^R} \\  
\end{array}
\right) =S_N^{-1}T_N^{tot}S_0\left( 
\begin{array}{c}
{\bf A^L} \\ {\bf B^L} \\  
\end{array}
\right) =M\left( 
\begin{array}{c}
{\bf A^L} \\ {\bf B^L} \\  
\end{array}
\right) 
\end{equation}
where the matrix $S$ is the transformation from the real space to the
momentum space, i.e. it provides the expansion in momentum components.

In order to determine the conductance, one needs to express the flux arising
along the lattice. This requires a further similarity transformation 
\begin{equation}
\label{trvel}F=\Gamma M\Gamma ^{-1}\;, 
\end{equation}
where the diagonal $2b\times 2b$ matrix $\Gamma $ is defined as $%
\Gamma_{i,j}=\delta _{i,j}\sqrt{v_i}$ for $i\le b$ and $\Gamma
_{i,j}=\delta_{i,j} \sqrt{v_{i-b}}$ for $i>b$. The above transformation is
equivalent to the normalization of the scattering matrix and it takes into
account the fact that each open channel propagates with a different
velocity. From Eq.~(\ref{tran}) it is easily seen that the transformation (%
\ref{trvel}) corresponds to the change of variables ${\bf a_i^{L,R}}= {\sqrt{%
v(p_i)}}{\bf A_i^{L,R}}$ and ${\bf b_i^{L,R}}={\sqrt{v(p_i)}} {\bf B_i^{L,R}}
$. The square of the modulus of the new variables gives the flux of
particles which enter and exit the disordered sample. It is easy now to
verify that $F$ conserves fluxes, 
\begin{equation}
\label{flux}F^{\dagger }\sigma _3F=\sigma _3 
\end{equation}
where $\sigma _3$ is a generalized Pauli $\sigma _z$ matrix, i.e. 
\begin{equation}
\sigma _3=\pmatrix{  1  & 0 \cr 0 & -1 \cr} 
\end{equation}
and 1 denotes a $b\times b$ identity matrix. Condition (\ref{flux}) is
crucial, since it corresponds to unitarity of scattering matrix.

By using the symplecticity of $F$ and $F^{\dagger }F$, and the relations
between $F$ and the scattering matrix, we find that the dimensionless
conductance (measured in $e^2/h$ units) is readily written as \cite{P84} 
\begin{equation}
\label{con} G = Tr \left (\frac{2}{F^{\dagger}F+(F^{\dagger}F)^{-1}+2}
\right) = \sum_{i=1}^b\frac 2{1+\cosh (2N\gamma _i(b,N))} 
\end{equation}
where the $\gamma _i(b,N)$'s are the Lyapunov exponents of the matrix $%
F^{\dagger}F$.

The lengths $\gamma _i^{-1}(b,N)$ depend both on $N$ and on the realization
of the disorder, but the ergodic multiplicative theorem \cite{O68} ensures
that, in the limit $N\rightarrow \infty $, $\gamma_i^{-1}(b,N)$ converges
towards the ``localization length'' $l_i^\infty $ 
\begin{equation}
\label{loc}\lim _{N\rightarrow \infty }\gamma _i^{-1}(b,N)=l_i^\infty
,\,\,\,\, 
\end{equation}
i.e. $l_i^\infty $ is a self-averaging quantity which depends on the energy $%
E$ and on the bandwidth $b$ only. Moreover, still in the limit $N\to \infty $%
, the contribution of the similarity transformations introduced to pass from
the transfer matrix $T_N$ to $F$ becomes negligible, implying that $%
\gamma_i(b,N) $ are equal to the logarithms of the eigenvalues of the matrix 
\begin{equation}
\label{alyap} O=\lim _{L\rightarrow \infty
}[(T_N)^{\dagger}(T_N)]^{1/2(L+2b)}\quad , 
\end{equation}
i.e. to the Lyapunov exponents of the product of the bare transfer matrices.

In the following we shall denote with $\gamma _1(b,N)> \gamma _2(b,N)>\ldots
>\gamma _i(b,N) >\ldots > \gamma _{2b}(b,N)$ the effective
(``finite-length'') Lyapunov exponents \cite{PV87} computed over a number $L$
of iterations at energy $E=0$. Because of the symplectic structure of $F$,
the Lyapunov exponents $\gamma _i(b,N)$ come in pairs with opposite values:
for this reason, it is sufficient to compute only the positive exponents.

\section{Lyapunov exponents}

The standard Landauer approach to quantum transport in a two-probe geometry
allows us to express the conductance $G$ in terms of $b$ real positive
Lyapunov exponents $\gamma _i$, see Eq.~(\ref{con}). Thus, the knowledge of
the joint probability distribution ${\cal P}(\{\gamma_i\})$ provides a
complete statistical description of the conductance. Accordingly, we start
this section by discussing the fluctuations of the Lyapunov spectra.

In the metallic regime, an expression for ${\cal P}(\{\gamma_i\})$ has been
proposed on the basis of Random Matrix Theory \cite{I86,PZIS90,P91,B94}.
Namely, it was conjectured that the matrix 
\begin{equation}
\label{pimatrix} X=\frac{F^{\dagger }F+(F^{\dagger }F)^{-1}-2}{4} 
\end{equation}
has the typical structure of full random matrices. This conjecture is based
on the observation that $F$ results from the product of many independent
random matrices so that its entries are random numbers all of the same
order. According to the RMT, the joint probability distribution of the
Lyapunov exponents {$\gamma _i$} is of the type 
\begin{equation}
\label{joint1} {\cal P}(\gamma _1,\gamma _2,\ldots ,\gamma _b) = C_\beta \;
\exp[-\beta H_N(\gamma _1,\gamma _2,\ldots ,\gamma _b)] 
\end{equation}
where $C_\beta $ is a normalization constant, $\beta =1, 2$ or $4$ is a
symmetry parameter and $H_N$ is the effective Hamiltonian 
\begin{equation}
\label{joint2} H_N(\gamma _1,\gamma _2,\ldots ,\gamma_b)=
-\sum_{i<j}^bu(\gamma _i,\gamma _j)+\sum_{i=1}^bV(\gamma _i). 
\end{equation}
Eq.~(\ref{joint1}) has the form of a Gibbs distribution at temperature $%
T=1/\beta$ for a fictitious system of classical particles on the semi
straight-line $(0,\infty)$ under the external potential $V(\gamma _i)$.
As for the two-particle interaction, in conventional RMT it has the
form
\begin{equation}
\label{joint3} u(\gamma
_i,\gamma_j)=-\ln|\cosh(2N\gamma_i)-\cosh(2N\gamma_j)|. 
\end{equation}
which corresponds to the logariphmic repulsion between the eigenvalues
$\lambda_i = cosh(2N\gamma_i)$.
All microscopic parameters are contained in the function $V(\gamma)$, while
the interaction potential $u$ is independent of them and has a geometrical
origin (see the discussion in \cite{B97}). The first impression was that, if 
$V(\gamma)$ is suitably chosen, the distribution (\ref{joint1},\ref{joint2}) 
with (\ref{joint3}) provides an
accurate description of transport properties in the metallic regime.
However, recent developments \cite{B94} have shown that the expression 
(\ref {joint3}) does not give the right value for the universal conductance
fluctuations in the quasi-1D limit. The correct result is obtained if the
two-body interaction potential (\ref{joint3}) is replaced by \cite{BR94}
(see details in \cite{B97}), 
\begin{equation}
\label{joint4} u(\gamma _i,\gamma _j)=-\frac 12\ln|\cosh(2N\gamma_i)-
\cosh(2N\gamma _j)|-\frac 12\ln|\gamma _i^2-\gamma _j^2|-\ln(2N) \quad . 
\end{equation}
Although the above expression has been derived for the case $\beta =2$ only,
i.e. for broken time-reversal symmetry, there are indications that it is
also valid for $\beta =1,4$.

Less is known about the properties of Lyapunov exponents in the localized
regime and in the metal-insulator transition (see \cite{M95} for a brief
review).

Below, we study numerically fluctuation properties of the Lyapunov exponents
$\gamma_1$.  
Standard techniques for the direct computation of eigenvalues are affected
by large inaccuracies due to the high probability of small denominators in
Eq.~(\ref{eq:mmap}). In order to avoid this problem, we have applied the
algorithm originally developed in \cite{BGGS80} for the infinite $N-$limit.
The method consists in considering a formally infinite sequence of matrices
all equal to $F^\dagger F$, which can be seen as ``replicas'' of the same
disordered sample. The Lyapunov exponents of the corresponding product of
matrices are then computed in a standard way by recursively applying the
single matrices to $b$ independent vectors which are continuously
orthonormalized (see also \cite{MK93,KPI97}). The advantage of this approach
is that one can keep the accuracy under control by renormalizing the vectors
also in the intermediate steps that correspond to the application of the
various matrices composing $F$.

In practice the number of ``replicas'' has been fixed by imposing that the
accuracy on the Lyapunov exponents is better than $10^{-4}$ for each
disorder realization. Moreover, ensemble averages have been performed over
sets of more than 2500 realizations.

In the metallic regime, the RMT predicts that level spacings 
for eigenvalues $\lambda_i$, and, therefore, the normalized differences 
\begin{equation}
\label{lyap1} \delta _i=\frac{\gamma _{i+1}-\gamma _i}{<\gamma_{i+1}-%
\gamma_i>}\quad ; \quad 1\le i\le b\quad ;\ \quad \gamma _{b+1}=0 
\end{equation}
of the Lyapunov exponents are distributed according to the
Wigner-Dyson (WD) surmise\cite{PZIS90,P91}. This expectation has been
numerically confirmed in the two- \cite{PZIS90} and three-dimensional \cite
{MK93} Anderson model.

In order to characterize the repulsion between any two consecutive Lyapunov
exponents, it is useful to introduce the normalized width $y$ of the
``level'' spacing distribution \cite{MK93}, 
\begin{equation}
\label{y}y(\delta _i)\equiv \frac{\sigma (\delta _i)}{<\delta _i>}\quad 
\end{equation}
where $\sigma$ denotes the corresponding standard deviation. For the
WD-distribution, the parameter $y$ is equal to $y_{WD}=\sqrt{4/\pi -1}\simeq
0.523$. Values of $y<y_{WD}$ are interpreted as the signature of a stronger
rigidity in the Lyapunov spectra. In particular, the limit $y\rightarrow 0$
corresponds to a $\delta $-distribution, i.e. to a ``crystallization'' of
the spectrum \cite{B97}. For small values of $y$, the distribution of the
spacings $\delta _i$ around their mean values is quite close to Gaussian;
this situation occurs in the localized regime when the repulsion is very
strong (see below). Notice that in the complete absence of repulsion (which
corresponds to a Poisson distribution), $y$ is equal to $1$ which thus
represents an upper bound.

It is known that the degree of localization in finite samples can be
effectively described by means of the scaling parameter 
\begin{equation}
\Lambda =b^2/N 
\end{equation}
which is connected to the $b^2$ proportionality of the localization length
in infinite samples (see, e.g., \cite{FM94}\cite{CMI90}). Numerical results
for three different values of the scaling parameter $\Lambda
=10.29,~6.95,~6.27$ in the metallic regime are presented in Fig.~1 by open
circles, diamonds and stars, respectively. The repulsion parameter $y$ is
reported versus the quantity $\chi = i/b$, the standard parameter used in
the representation of the whole Lyapunov spectra (see also \cite
{KPIR96,KPI97}). As one can see, the value of $y$ is almost constant for $%
\chi$ larger than 0.2. Although, in average, it is very close to the
theoretical prediction $y_{BW}$, there is a systematic difference, namely, $%
y>y_{BW}$. This is a clear indication of a disagreement with respect to RMT
expectations. The only exception is a narrow region around $\chi = 1$ where $%
y$ is slightly smaller and thus more in agreement with the RMT.

Larger Lyapunov exponents, in the region $\chi \leq 0.2$, are characterized
by a much larger deviation from the WD-value. In particular, for $\chi
\rightarrow 0$, the value of $y$ decreases, revealing a much stronger
repulsion than predicted by the Random Matrix Theory. This result is in
contrast to previous studies for the 2D Anderson model \cite{MK93,PZIS90}
where it was found that the distribution for $\delta _i$ follows the WD-form
for all Lyapunov exponents.

Now, we analyze the statistical properties of the lowest Lyapunov exponent $%
\gamma_b(b,N)$ for different values of the scaling parameter $\Lambda$.
Numerical data shows that the distribution of $\gamma _b$ changes
dramatically when $\Lambda $ decreases. For example, for $\Lambda \approx 0.6
$, the distribution ${\cal P}(\delta _b)$ turns out to be very different
from the WD-form, and for $\Lambda \approx 0.2$ it looks like the Gaussian
distribution around the mean value of $\delta _b$. In order to describe this
transition quantitatively, we have fitted numerical data to the expression 
\begin{equation}
\label{felixdi} {\cal P}(\delta )=A\delta ^\beta (1+B\beta \delta
)^{f(\beta)} \exp\left( -\frac{\pi ^2}{16}\beta \delta ^2-\frac \pi
2(1-\beta /2)\delta \right) 
\end{equation}
where $f(\beta)$ is the fitting function, $f(\beta )=2^\beta (1-\beta
/2)/\beta -0.16874$. The constants $A$ and $B$ are defined through the
normalization conditions 
\begin{equation}
\label{norm} \int {\cal P}(s)\;ds=1;\;\;\;\;\int s\; {\cal P}(s)\;ds=1\;\;\; 
\end{equation}
while the free parameter $\beta$ is the effective repulsion taking values in 
$[0,\infty)$. The above phenomenological expression has been suggested in
Ref. \cite{CIM91} (see \cite{I93} for details) in order to describe the
level spacing distribution for quasi-1D disordered samples of finite size,
in the dependence on the degree of localization. A similar distribution has
been also used in the description of the spacing between real particles (on
a ring) interacting with each other via 2D Coulomb forces. In that case, $%
\beta$ plays the role of the inverse temperature in the thermodynamic
equilibrium \cite{IS89,I90}. For $\beta =0$, the dependence ${\cal P}(\delta)
$ reduces to the Poisson distribution, while for $\beta =\infty $ it
converges to a $\delta$-function centered at $\delta=1$. Moreover, for $%
\beta = 1,2, 4$ the expression (\ref{felixdi}) is very close to that one
given by the RMT (it even turns out to be more accurate than the WD-surmise,
see \cite{I93}).

The ``inverse temperature'' $\beta $ is a very appropriate indicator to
characterize the transition from a totally uncorrelated Lyapunov spectrum ($%
\beta =0$) to a ``perfect crystal'' ($\beta =\infty $). In fact, it can be
effectively determined by fitting the data with expression (\ref{felixdi}).
In order to emphasize the validity of our results, let us notice that the
quality of the fit has been quantitatively checked by performing a $\chi^2$
test, that has been always brilliantly passed. In the metallic regime (e.g.,
for $\Lambda =10.24$), the best fit of the numerical data for $\gamma _b$
gives $\beta =1$ with a high accuracy, indicating that the minimum Lyapunov
exponent is, indeed, characterized by the WD-distribution \cite{P91}. When
moving towards the localized regime, where $\Lambda \ll 1$, the value of the
repulsion parameter increases; for $\Lambda \approx 0.2$, we have $\beta
\approx 3.95$. One can study the transition from one to the other regime by
computing the repulsion parameter $\beta $ for different values of $b$ and $N
$. The good data collapse observed in Fig.~2 strongly indicate that $\beta $
(for the minimum Lyapunov exponent) is a function of the scaling parameter $%
\Lambda =b^2/N$  only, a result which is in the spirit of previous studies 
\cite{CMI90,FM94}. Moreover, one can observe that already for $\Lambda \ge 4$%
, the repulsion parameter reaches the lower limit $\beta =1$ (we remind
that, being the condition $b\ll N$ fulfilled, we are, for $\Lambda \gg 1$,
in the metallic regime). For $\Lambda <1$, the repulsion parameter starts
increasing very fast and in the limit $\Lambda \rightarrow 0$ it diverges.
In other words, in the extremely localized regime, the probability
distribution for the lowest Lyapunov exponent is a $\delta $-like function,
which can be approximated by a Gaussian. A similar effect has been observed
in previous studies of 2D Anderson-type models \cite{MK93,MK93b}.

Finally, for what concerns the properties of the other Lyapunov exponents in
the localized regime, we can look at full circles, diamonds and triangles in
Fig.~1. First, we can notice that for all $i$'s, including $i=b$, the
quantity $y(\delta_i)$ decreases as we move towards the localized regime,
leading to a Gaussian-like distribution. This is in agreement with the
previous results for the minimum Lyapunov exponent $\gamma _b$ for which the
same conclusion was reached by measuring directly the repulsion parameter $%
\beta$. Moreover, in comparison with the metallic regime, the convergence to
the $\delta$-like distribution for the maximal part of the spectra becomes
smoother. One should note that in the limit $N\rightarrow \infty$, the
distribution $P(\delta_i)$ for $i<b$ was instead found to be Poissonian in
the localized regime of the 2D Anderson model \cite{MK93,MK93b}.  This means
that the quasi-1D limit which we consider here, is somewhat special. Namely,
the eigenvalues of the transfer matrix $F^{\dagger }F$ are well separated,
their number $b$ being fixed while their mean separation increases linearly
with $N$. On the other hand, if we consider a two-dimensional square of size 
$N\times N$, then $b$ will also diverge with $N$ and the assumption of well
separated eigenvalues is no longer valid.

\section{Conductance}

The present understanding of Anderson localization theory is mainly based on
the one-parameter scaling. However, recent experimental results indicate
that in order to understand the transport properties of disordered systems,
one should study not only the mean expectation values but also the whole
probability distribution.

One of the quantities that attracted a lot of research interests during last
two decades, is the conductance. Theoretical and experimental results have
shown (see references in \cite{B97}) that the variance of the conductance is 
${\cal O}(1)$ (in $e^2/h$ units), independently of the average conductance,
if the conductor is phase-coherent. A number of theoretical methods has been
developed to compute the (universal) conductance fluctuations arising in the
above mentioned (metallic) regime. Many numerical calculations have also
been performed although, until recently, mostly in connection with the
Thouless approach which assumes that the system is closed (on this subject,
see \cite{BHMM96}). On the other hand, very little efforts have been made to
study open systems, taking into account the specific form of the coupling
with the leads \cite{CGM97,M97}, as in our case. Yet, it is known that the
conductance of a mesoscopic sample depends sensitively on the measurement
geometry and on the way the leads are attached to the sample \cite{ES81}. In
this section we study the properties of the conductance $G$ in the quasi-1D
model (\ref{eqmo}) and compare them with theoretical expectations and
numerical results obtained in a slightly different model \cite{CGM97}.

Our first interest is in the geometric average of the conductance $\langle
\ln G \rangle$ in the metallic regime $1\ll b\ll N$ (here, the brackets $%
\langle \ldots \rangle$ denote an average over different realizations of the
random potential). As expected, the average conductance $\tilde G \equiv
\exp\left (\langle \ln G\rangle\right)$ turns out to depend on $b$ and $N$
only through the scaling parameter $\Lambda =b^2/N$ (see Fig.~3, where we
have reported the outcome of several simulations performed for different
values of $N$ and $b$). The existence of a specific function $\tilde G
(\Lambda)$ is equivalent to the standard assumption made in the scaling
theory \cite{AALR79,book}, i.e. to assuming that the logarithmic derivative
of the conductance, 
\begin{equation}
\label{bfun} \eta \equiv \frac{d \ln \tilde G}{d\ln N} 
\end{equation}
is a function only of $\tilde G$. The representation of the numerical
results in terms of $\eta$ and $\tilde G$ is given in Fig.~4.

From the inset of Fig.~3, it can be seen that in the metallic regime, i.e.
for $\Lambda \gg 1$, $\tilde G\simeq a_1\Lambda +a_0$, that is the
conductance has the expected ``ohmic'' dependence on the sample length ($%
\Lambda \sim 1/N$). In the $\eta $ representation, this is tantamount to
saying that $\eta \simeq -1+a_0/{\tilde G}$ with $a_0<0$. In the localized
regime, $\Lambda \ll 1$, Fig.~3 reveals an exponential decrease of the
conductance with the chain length, $\tilde G\simeq b_0\exp (-b_1/\Lambda)$,
a behavior which corresponds to $\eta =\ln ({\tilde G}/b_0)$.

Let us now discuss some approximate theoretical expressions of $G$ and
compare them with our numerical results. First of all, we want to derive a
sufficiently accurate expression for the effective Lyapunov exponents in
order to determine the conductance from the Landauer formula in the limit of
large $b$ and $N$ but finite $\Lambda$. The extensive studies performed in
Ref.~\cite{KPI97} have suggested that 
\begin{equation}
\label{gfin1} \gamma_i(b,N) \approx \gamma_i(b,\infty) - \frac{\Lambda}{%
b^{1.7+i/b}} \quad , 
\end{equation}
is an expression convincingly tested in the bulk of the spectrum. Moreover,
it was shown that 
\begin{equation}
\label{gfin2} \gamma_i(b,\infty) \approx \frac{\Omega(\chi)}{b} + \frac{%
\omega(\chi)}{b^2} 
\end{equation}
where $\Omega(\chi)$ represents the asymptotic Lyapunov spectrum, while $%
\omega(\chi)$ is the leading correction term. By substituting Eq.~(\ref
{gfin2}) in Eq.~(\ref{gfin1}), one obtains 
\begin{equation}
\label{gfin3} \gamma_i(b,N) \approx \frac{\Omega(\chi)}{b} + \frac{%
\omega(\chi)}{b^2} - \frac{\Lambda}{b^{1.7+\chi}} 
\end{equation}
As the leading contribution to the conductance is given by the ``small''
Lyapunov exponents (i.e. $\chi \simeq 1$), it is convenient to expand $%
\Omega(\chi)$ around $\chi=1$. By retaining the leading terms (recall that $%
\Omega(1) = 0$) one obtains 
\begin{equation}
\label{gfin4} \gamma_i(b,N) \approx \Omega_1 \frac{(b-i)}{b^2} + \Omega_2 
\frac{(b-i)^2}{b^3} + \frac{\omega(1)}{b^2} - \frac{\Lambda}{b^{2.7}} 
\end{equation}
where $\Omega_1$ is the absolute value of the first of derivative of $%
\Omega(\chi)$, while $\Omega_2$ is the second derivative.

From the Landauer expression (\ref{con}), one can see that the argument of
the hyperbolic cosines is 
\begin{equation}
\label{gfin5}\alpha _j\equiv 2\gamma _j(b,N)N\approx \frac{2\Omega _1}%
\Lambda j+\frac{2\Omega _2}{b\Lambda }(j-1)^2+2\frac{\omega (1)-\Omega _1}%
\Lambda -\frac{2\Lambda }{b^{0.7}}
\end{equation}
where $j=b-i+1$ Accordingly, one can see that in the limit $b$, $N\to \infty 
$ but finite $\Lambda $, the second and the last term in the r.h.s. vanish
and thus we neglect them.

Therefore, we are left with a linear expression in $j$, with two
coefficients to be determined, namely $\Omega_1$ and $\omega(1)$. From the
analytic knowledge of the minimum Lyapunov exponent ($j=1$), we have that $%
\omega(1) = 3/2$. Moreover, previous simulations for the Lyapunov spectra
have clearly indicated that the slope is $\Omega_1 = 1.5$. As a consequence,
we are left with 
\begin{equation}
\label{gfin6} \alpha_j = 3 j/\Lambda \quad , 
\end{equation}
which, upon substitution into Eq.~(\ref{con}) implies 
\begin{equation}
\label{conap1} G=\sum_{j=1}^\infty \frac 2{1+\cosh (3j/\Lambda )}, 
\end{equation}
where, in view of the limit $b \to \infty$, the sum is extended to infinity.

In the metallic regime, the conductance expression can be written in a
compact form by transforming the sum into an integral, 
\begin{equation}
\label{conap2} G=\frac{2\Lambda }3\int_{3/\Lambda }^\infty {\frac{dy}{%
1+\cosh y}}, 
\end{equation}
where we have introduced the variable $y=3j/\Lambda $. The solution of the
integral finally yields, 
\begin{equation}
\label{conap3} G=\frac{4\Lambda }{3(\exp (3/\Lambda )+1)}. 
\end{equation}
In the limit $\Lambda \gg 1$ one gets 
\begin{equation}
\label{metall} G=a_1\Lambda \;\;+\;a_0\;=\;\frac {l_b^\infty}{N}\;-\;1 
\end{equation}
with $a_1=2/3$ and $a_0=-1$ . The first term in (\ref{metall}) coincides
with the theoretical result according to which the leading term for the
conductance in the metallic regime is the ratio of the localization length $%
l_b^\infty$ to the size $N$ of the sample. One should stress that the finite
size corrections are neglected in this term (see, e.g., \cite{B97}). A more
important correction is given by the term $a_0$ and is related to weak
localization effects. The meaning of this quantum correction is that the
back scattering is larger in comparison to the classical result and,
therefore, the conductance is smaller than the contribution given by the
first term. In the theory (see the review \cite{B97} and references
therein), the value of $a_0$ is known to be $a_0=-1/3$ \cite{M88}. It is,
however, interesting to note that a simple estimate of this correction \cite
{D95} gives the value $a_0=-1$ as in our expression (\ref{metall}).

In the metallic regime, the best fit of our numerical data with the
expression (\ref{metall}) gives the values $a_1\approx 0.74$ and $a_0\approx
-0.41\;$. While the value of $a_1$ is close to the theoretical prediction,
the correction term $a_0$ is somehow different. A discussion about the
correspondence between numerics and theory is given below.

We now go to consider the opposite limit of strong localization, $\Lambda
\ll 1$. In this case, the sum (\ref{conap1}) is dominated by the first
contribution, 
\begin{equation}
\label{local} G\;=\;b_0\;\exp \left( -\frac{b_1}\Lambda \right) 
\end{equation}
with $b_0=2$ and $b_1=3$ . Let us first discuss the leading dependence given
by the term $b_1\,.$ One should first stress that the coefficient $b_1$ is
different for different definitions of the average. Specifically, if one
treat $G\;$ as the average conductance $\left\langle G\right\rangle $, the
value of $b_1$ was found to be $b_1=3/4$. For the geometric average $\tilde
G\;$, instead, (also known as the ``typical conductance''), the theoretical
prediction is $b_1=3$ (see \cite{Z92,B97}). Therefore, the expression (\ref
{local}), which refers to the latter case, gives the correct exponential
dependence.

Less is known about the correction to the exponential dependence which in (%
\ref{local}) is absorbed in the term $b_0\,.$ To the best of our knowledge,
this correction is unknown for $\tilde G\,,$ the quantity of our numerical
interest. At the same time, for the average conductance $\left\langle
G\right\rangle \,$the correction has the form $b_0\sim \Lambda ^{3/2}$ $\cite
{Z92}$ . One can suggest, however, that for the geometric conductance $%
\tilde G\,$ the correction has the form $b_0\sim \Lambda ^{\nu \text{ }}$
with some value $\nu \geq 1$. Our data do not allow to estimate the
correction to the exponential dependence. The best fit to the expression (%
\ref{local}) with $b_1$ as a constant, gives the following values, $%
b_1\approx 2.9$ and $b_0\approx 5.58$. One can see that the value of $b_1\,$
is quite close to the theoretical one, $b_1=3$.

A comparison of the theoretical expression (\ref{conap1}) with the data in a
broad interval of the localization parameter $\Lambda $ is presented in
Fig.~3. Considering that no free parameters are present in the theoretical
formula, the overall agreement represents a nice confirmation of the various
approximations behind Eq.~(\ref{conap1}). Nonetheless a small but clear
deviation can be seen in the intermediate region $\Lambda \sim 1$. It is,
therefore, natural and important to understand whether the disagreement is
to be ascribed to the misjudgement of some correction term, or to finiteness
of $b$ and $N$ in the simulations. Moreover, it is important to recall that
conductance fluctuations are not taken into account in the derivation of
Eq.~(\ref{conap1}), since it is expressed in terms of the mean Lyapunov
exponents. This analysis goes beyond the scope of the present paper.

One can also describe our data for the conductance $\tilde G$ by making use
the phenomenological expression suggested in \cite{CGM97}, 
\begin{equation}
\label{scon}\tilde G\;\equiv \exp \;\langle \ln G\rangle \;=\;\frac{%
c_1+c_2\exp (-c_4/\Lambda )}{\exp (c_3/\Lambda )-1}.
\end{equation}
with four parameters $c_i$. By fitting directly all the parameters, one
finds an overall excellent correspondence with our data, but a careful
inspection of the behavior in the metallic and localized regimes is less
convincing. On the other hand, one can determine the parameters $c_i$, in
such limit cases by exploiting the correspondence between $c_i$ and $%
a_0\;,\;a_1\;,\;b_0$ ,\ $b_1$ , see Eqs. (\ref{metall}) and (\ref{local}), 
\begin{equation}
\label{abc}a_1=\frac{c_1+c_2}{c_3}\;;\;a_0=-\frac{c_2c_4}{c_3}-\frac{c_1+c_2}%
2\;;\;b_1=c_3\;;\;b_0=c_1
\end{equation}

Unexpectedly, the resulting dependence (\ref{scon}) with these (different)
values of $c_i$, turns out to be practically indistinguishable from the
analytical expression (\ref{conap1}). Therefore, the dependence (\ref{scon})
has no advantage in comparison with Eq.~(\ref{conap1}). It should be noted
that in \cite{CGM97}, where the same problem has been studied with a
different method (scattering matrix approach) and different boundary
conditions (the coupling matrix elements have been chosen to be constant as
in the leads, instead of random as in the bulk), the parameters $c_i$ and
the behavior in the metallic regime are slightly different. At the moment,
we cannot give a convincing explanation of this discrepancy. In fact,
theoretically, the influence of the coupling is still not understood, while,
numerically, the quality of the data is not such to allow drawing a definite
about the influence of the coupling.

We have also performed a detailed analysis of the distribution ${\cal P}(G)$
in various regimes. In particular, it was found that the shape of the
distribution is controlled by the same scaling parameter $\Lambda$. In the
localized region, the distribution is approximately log-normal (see Fig.~5).
The variance of $\ln G$ decreases upon increasing $\Lambda$ i.e. moving
towards the metallic regime and in the localized regime satisfies the
following relation 
\begin{equation}
\label{varg} var(\ln G)=-A_0\ln G-B_0, 
\end{equation}
where $A_0\simeq 1.98\pm 0.03$ and $B_0\simeq 3\pm 0.4$ (see Fig.~6). The
value of the slope is fully consistent with theoretical expectations and
with the numerical results obtained in \cite{CGM97}. However, let us recall
that in the computation of Thouless conductance, it was found that $%
A_0\simeq 1$ \cite{CGIMZ94}. This means that fluctuations properties of
Landauer and Thouless conductance are different (see also \cite{BHMM96}).
The value of the constant $B_0$ is instead slightly different from what
found in \cite{CGM97}. This should be again a consequence of the different
connection between the disordered sample and the leads.

Moving towards the metallic regime, the shape of the distribution ${\cal P}%
(G)$ becomes normal (see Fig.~7). This result is in perfect agreement with
the theoretical predictions for quasi-1D conductors \cite{PZIS90}.

The last problem that we have addressed in this paper is related to the
so-called Universal Conductance Fluctuations. As is known (see, for example,
the review \cite{B97}), the correct value of the variance of the
dimensionless conductance is $var(G)=2/15$. At the same time, direct
application of the standard Random Matrix Theory to transmission matrices in
the metallic regime gives $var(G)=1/8$ \cite{B97}. The difference is small
but crucial, it indicates that the repulsion between the transmission
eigenvalues is not correctly described by the logarithmic two-body
interaction, which is the core of the Random Matrix Theory. In this sense,
it is important to study directly in numerical experiments how the
fluctuations of the conductance depend on the degree of localization. We
have performed simulations for different values of the scaling parameter $%
\Lambda $. Our data for the variance are reported in Fig.~8. As one can see,
for $\Lambda $ still approximately equal to $4.5$, the variance remains
close to $1/8$, the value predicted by the RMT. However, a further increase
of the scaling parameter up to $\Lambda \approx 7.4$ leads the fluctuations
to approach $2/15$, the value expected from either diagrammatic calculations
or the diffusion-equation approach for transfer matrices (for the references
see \cite{B97}). One should stress that it is important to satisfy the
condition $b\ll N$ in order to distinguish the metallic regime from the
ballistic one. The latter, highly non-generic, regime is reflected in the
points in Fig.~8 which deviate from above when this condition is violated.
The data clearly indicates that the variance reaches the limit value 2/15
very slowly, an effect revealed also by a different approach, based on the
scattering matrix determined through the numerical solution of the
Lippman-Schwinger equation, see \cite{CGM97}.

\section{\bf Discussion }

In this paper we have studied the statistical properties of the Lyapunov
spectra for quasi-1D disordered samples of finite size in the connection
with electric conductance and its fluctuations. The key point for our study
is represented by Landauer expression, giving the conductance in terms of
the Lyapunov exponents of a suitable transfer matrix. The mathematical model
we have used is the band matrix ensemble with random entries for the
disordered part.

The central question we have numerically investigated, is how statistical
properties of the Lyapunov exponents are reflected in the properties of the
conductance. The main theory which gives quite a good description of the
properties of the conductance, is the standard Random Matrix Theory (RMT)
applied to transfer matrices. The conjecture that the RMT is a good
mathematical tool for a theoretical approach, is based on the observation
that different channels are strongly but irregularly coupled. This approach
has led to many interesting theoretical results \cite{B97} which have found
quite a good support in several numerical experiments. Nonetheless, some
small deviations can be found with respect to the RMT predictions. The most
famous concerns the value of the normalized variance of the conductance
fluctuations which is $2/15$ instead of $1/8$. The difference is small but
important for the theoretical implications about the universal properties of
the conductance fluctuations.

The origin of the above discrepancy has been attributed to different
structure of the Lyapunov spectra compared to the spectrum of truly random
matrices. For instance, in the RMT, the eigenvalues occupy the infinite line
(for infinite random matrices with a fixed variance of the matrix elements),
while the Lyapunov spectrum of the product of transfer matrices is symmetric
with respect to the origin, a circumstance yielding an additional
(non-random) repulsion between negative and positive ``levels'', which is
absent in completely random matrices. On the other hand, in the diffusive
regime, given the large number of channels, such an effect may be neglected.
Indeed, earlier numerical experiments have shown quite a good agreement with
the RMT prediction; for example, the level spacing distribution was found to
be well described by the famous Wigner-Dyson (WD) surmise (for the lowest
Lyapunov exponent in the diffusive regime).

In our study, we have paid special attention to the whole Lyapunov spectrum,
performing some qualitative tests aimed at emphasizing deviations of the
level statistics from the Wigner-Dyson form. By making use of the normalized
variance of the level spacing distribution, we have found that the
statistical properties of the lowest Lyapunov exponent (as well as of the
second one) are, indeed, described very well by the WD form. However, we
have also detected a clear deviation from the RMT predictions for all the
other exponents. In particular, it was found that the repulsion is weaker
than that predicted by the RMT. This fact is quite unexpected since one
might conjecture that the strongest influence of the negative part of the
spectrum occurs for the lowest Lyapunov exponent, not for the bulk. Indeed,
the lowest Lyapunov exponent has a ``random'' repulsion from one side only;
on the negative side, the repulsion is of a different ``regular'' nature,
due to the symmetry. However, it is precisely the lowest Lyapunov exponent
to exhibit the best agreement with the random matrix approach (see also \cite
{P91}). Thus we are lead to conjecture that besides the symmetry, a further
feature must be present which contributes to differentiate a transfer matrix
from purely random one. Anyway, the deviations observed in the bulk of the
Lyapunov spectrum represent a further more detailed evidence of the failure
of the RMT which has to be added to the difference between $1/8$ and $2/15$
for the conductance fluctuation (see also discussion and references in \cite
{B97}).

To study the level spacing distribution in the general case (when the
distribution does not coincide with the WD form), we have used the
phenomenological expression suggested in \cite{CIM91} (see details in \cite
{I93}) for the description of the so-called intermediate statistics which
occurs due to strong localization effects. As long as the effective
repulsion ranges between $0$ (Poissonian) and $\beta=1$ (WD-distribution for
the Gaussian Orthogonal Ensemble, GOE), the expression could be compared
with the Brody-distribution \cite{B73}. However, in the present case, the
effective repulsion is larger ($\beta > 1$), even diverging in the limit of
strong localization, where the distribution becomes an increasingly narrow
Gaussian (a feature already discussed in the literature \cite{PZIS90}). Our
analytical expression for the level spacing distribution allows to extract
the repulsion parameter and to determine its dependence on the degree of
localization (see Fig.~2). In analogy to other applications \cite{I90,CIM91}%
, the effective repulsion turns out to depend only on the scaling parameter $%
b^2/N$.

In the second part of the paper, we have investigated various properties of
the conductance in a wide range of the localization parameter $\Lambda$. In
particular, we have numerically computed the dependence of the geometric
("typical") conductance $\tilde G$ on $\Lambda$, from the very localized to
the diffusive regime. One should recall theoretical results on the
dependence of the conductance on $\Lambda $ are available only in the limit
cases of a strong localization and in the metallic regime.

Starting from the Landauer expression and taking into account the main
properties of the Lyapunov spectra \cite{KPIR96,KPI97}, we have obtained an
analytical expression for the conductance with no free parameters.
Evaluation of this expression in the two opposite (metallic and localized)
limits have shown that this expression gives, in general, the correct
behavior in comparison with some theoretical predictions. In particular, in
the metallic regime the leading term is exactly the same as predicted by the
theory, although the weak localization correction is slightly different.

A clear (but not very strong) difference has been found in the intermediate
region $\Lambda \approx 1$. The source of this discrepancy is still not
clear, in any case, one should emphasize that in this region, both the
coupling to the leads and finite size effects can play a significant role
(we would like also to stress that there is no analytical prediction in this
region for $\tilde G$). Moreover, we would like to note that the question of
the influence of the coupling on the statistical properties of the
conductance is not theoretically understood even for the relatively simple
case of quasi-1D geometry. Recent numerical experiments \cite{GIM97} with
the same model (\ref{eqmo}) have shown that by changing the degree of
coupling, the structure of scattering states changes dramatically. In our
present study, we have used the so-called ``matching'' principle according
to which the variance of coupling elements is the same as that one in the
bulk (see also \cite{KPI97}); in this case the influence of the coupling is
expected to be negligible theoretical predictions can be used.

Finally, our results for the conductance fluctuations are in excellent
agreement with the theoretical predictions. In particular, the distribution
of the conductance has been found to be log-normal in the localized regime
and of Gaussian in the diffusive regime. Furthermore, in the localized
regime, the proportionality coefficient between the variance of the
logarithm of the conductance and the logarithm of the conductance itself, is
very close to the theoretical value 2. It is interesting to note, that when
determining the conductance according to the Thouless definition via the
curvature of the levels of a closed systems, this value was instead found to
be 1 \cite{CGIMZ94}. This fact indicates that fluctuation properties of the
curvature in the localized regime cannot be associated with that of
conductance fluctuations (see also \cite{BHMM96}).

In the metallic regime, we have carefully studied the universal conductance
fluctuations. One should stress that, numerically, this problem is far from
trivial, since the sample length $N$ has to be larger than band size $b$
(defining the mean free path) but smaller that $b^2$ ($N\ll b^2$, where $b^2$
is a measure of the localization length in infinite samples). On the other
hand, one also needs a sufficient statistics in order to discriminate
between $1/8$ and $2/15$ for the value of the variance of the normalized
conductance. We have performed an extensive numerical analysis to overcome
the above mentioned computational difficulties. The main result we have
obtained is the observation of a slow convergence to the theoretical value $%
2/15$ upon increasing the localization parameter $\Lambda $. Nonetheless, we
have also found a ``pre-diffusive'' regime, $\Lambda \approx 4.5$ where the
variance stays very close to $1/8$. This is an indirect indication that many
Lyapunov exponents come into play, besides the lowest one, which have
fluctuations slightly different from that given by conventional RMT. Thus,
both the universal conductance fluctuations and the origin of the difference
between $1/8$ and $2/15$, predicted theoretically in \cite{B94}, are nicely
confirmed by our numerical study of the quasi-1D model.

\acknowledgments

We would like to acknowledge G.~Casati, I.~Guarneri and L.~Molinari for useful
discussions on scattering problems. One of us (T.K.) would like to thank the
Istituto Nazionale di Ottica for the hospitality during the fall of 1995 and
to acknowledge the support of Grant CHRX-CT93-0107. 
F.M.I. acknowledges the support received from
 the Cariplo Foundation for Scientific Research as well
 as the partial support by the Grant No. INTAS-94-2058.

%\newpage

\begin{figure} 
\caption{Parameter $y$ (see Eq.~(\ref{y})) for different
values of the scaling parameter $\Lambda=b^2/N$ versus the
normalized number $\chi = i/b$ of the Lyapunov exponent. 
The horizontal line corresponds to the Wigner-Dyson surmise.}
\end{figure}

\begin{figure} 
\caption{Effective repulsion parameter $\beta $ is plotted 
versus $\Lambda$ for the minimum Lyapunov exponent $\gamma_b$.}
\end{figure}

\begin{figure} 
\caption{Average logarithm of the conductance versus the logarithm of
the scaling parameter $\Lambda$ for different values of the parameters 
$N,b$ with $N=L+2b$. The solid curve is the expression (31).
In the inset the geometrical average 
of conductance versus $\Lambda$ is given for the metallic regime 
$(l_{\infty} \sim b^2 \gg N)$.  
The best least square fit gives $y=ax+b$ with $a\simeq 0.75$ and
$b\simeq -0.6$.}
\end{figure}

\begin{figure}
\caption{Scaling function $\eta$. The horizontal line gives
the asymptotic limit $\eta = 1$ for $G \gg 1$. 
Another line is used as a guide for the 
asymptotic slope of $ \eta $ in the localized regime. 
In the inset we present the data for $\eta $ in the $G>>1$ limit.
Straight line corresponds to the best least square fit (see in
the text).}
\end{figure}

\begin{figure}
\caption{Log-normal distribution of the conductance in the localized
regime.}
\end{figure}

\begin{figure}
\caption{Variance of the logarithm of the conductance against
the 
average logarithm of the conductance in the localized regime 
$\Lambda \ll 1$. The fit
shows that the variance of the logarithm is two times larger
than the mean value
of the logarithm of conductance.}
\end{figure}

\begin{figure}
\caption{Normal distribution of the conductance in the metallic regime
$\Lambda \gg 1$.}
\end{figure}

\begin{figure}
\caption{Variance of the logarithm of conductance versus the logarithm of the
scaling parameter $\Lambda$. The low straight line is 1/8, the upper is 2/15.
The error bars are of the order of the symbols. For all cases, we have
averaged over more than 2500 different realizations of the disorder.}
\end{figure}

\end{document}